# Coherent Dynamics of Floquet-Bloch States in Monolayer WS$_2$ Reveals Fast Adiabatic Switching

S.K. Earl[1,2], M.A. Conway[1,2], J.B Muir[1,2], M. Wurdack[3], E.A. Ostrovskaya[3], J.O. Tollerud[1,2] & J.A. Davis[1,2]

[1]*Optical Sciences Centre, Swinburne University of Technology, Victoria 3122, Australia*

[2]*ARC Centre of Excellence in Future Low-Energy Electronics Technologies, Swinburne University of Technology, Victoria, 3122, Australia*

[3]*ARC Centre of Excellence in Future Low-Energy Electronics Technologies and Nonlinear Physics Centre, Research School of Physics, The Australian National University, Canberra, ACT, 2601, Australia*

jdavis@swin.edu.au

**Abstract:**

Floquet engineering offers a path to optically-controlled materials, but experimental implementations have frequently relied on femtosecond pulses to achieve the high peak fields required to maximise interactions and obtain a measurable response. The Floquet formalism, however, is based on a continuous, periodic drive. Here we use femtosecond laser pulses to drive the optical Stark effect, a simple realization of Floquet engineering, in monolayer WS$_2$. By monitoring the coherent evolution of the free-induction decay, we show that the system evolves adiabatically, and that the finite duration of the pulses does not introduce any effects beyond Floquet theory. Furthermore, we demonstrate that the induced energy shift follows a linear dependence on instantaneous intensity, even for ultrafast driving fields of fewer than 15 optical cycles.

Non-equilibrium phases of matter can provide access to microscopic physics, and macroscopic properties, unavailable in extant materials[1–5]. Explorations of materials driven out of equilibrium has led to the discovery of a range of phenomena such as a light-driven anomalous Hall effect[6,7], light-induced superconductivity[8,9] and light-induced phase transitions[10,11]. Floquet engineering[12], the process of using a periodic perturbation to reversibly manipulate bandstructure, is one avenue to control material properties. Floquet theory removes the time-dependence of a Hamiltonian, resulting in an infinite, periodic set of replica bands spaced by the frequency of the perturbation, known as Floquet sidebands[13]. The electromagnetic field oscillations of light can act as the periodic perturbation while careful selection of the energy and other properties can be used to manipulate interactions between the equilibrium bands and their Floquet replicas to control the resulting non-equilibrium properties. Floquet-Bloch bands generated in this manner have been observed, for example, as replicas of the surface states of a topological insulator in time- and angle-resolved photoemission spectroscopy (tr-ARPES) measurements[13,14].

The application of Floquet theory to both the Haldane model[15] and HgTe/CdTe quantum wells[16] predicted a reversible change in the topology of the bandstructure in both cases. Cold atom experiments have been used to simulate these predictions, showing changes to the topological phase via the Chern number[17,18], and the emergence of chiral edge states[19]. Additionally, experiments with graphene and a circularly polarized mid-IR driving field appear to show the realization of a topologically non-trivial band structure and the opening of a gap, as evidenced by an anomalous Hall effect[6,7,20]. A topological phase transition is also predicted to arise in semiconducting monolayer transition metal dichalcogenides (TMDs) when driven with an optical field with frequency larger than the band gap[21,22], but has yet to be observed experimentally.

In order to elicit a response at a detectable level, ultrafast laser pulses with high peak intensities are commonly used in these experiments because the interactions between states typically scales with the driving field amplitude. Floquet theory, however, is predicated on a continuous, monochromatic drive[12,16]. The discrete nature of these pulses then raises new questions: How short can pulses be before Floquet theory ceases to be applicable? When does the bandstructure (i.e., the eigenstates of the system) cease to evolve adiabatically?

The optical Stark effect (equivalently, the AC Stark effect) is a well-understood phenomenon[23] that emerges naturally from Floquet theory[24]. It provides an ideal model to test the limits of the Floquet formalism for short pulses and the adiabaticity of turning on the periodic driving field on femtosecond timescales. In TMD monolayers[21,22,25–28], a valley-selective optical Stark shift has been observed, with the magnitude of the shift obeying the expected linear dependence on pump intensity, and inverse dependence on the detuning of the pump from the transition energy[29]. Coherence between valleys was shown to be maintained to some extent, though not quantified, and the phase rotated by the shift[25]. However, little consideration has been given to the duration of the driving field, with the majority of these measurements using pulses longer than 100fs. Experiments investigating the optical Stark effect in GaAs quantum wells/bulk GaAs found that the adiabatic approximation holds with a >100 fs sub-bandgap pump pulse[30–32]. A Green's function analysis, in agreement, found that the width of the Gaussian envelope was the best predictor of convergence to the case of a continuous drive[33], and Floquet analysis applied to the turning on of a pulse showed a similar result[34].

Here, we drive the optical Stark effect in monolayer $WS_2$ with red-detuned pump pulses as short as 34 fs (~15 optical cycles) and probe the response with 25 fs pulses, resonant with the exciton transition. By varying the pump pulse duration while maintaining constant pulse energy, we show that the observed shift scales linearly with peak intensity, as predicted by Floquet theory, and, more significantly, follows (within our time resolution) the instantaneous intensity of the pulse envelope, showing no deviation from Floquet theory and the adiabatic approximation. Furthermore, we observe coherent dynamics of the free induction decay, revealing a dephasing time of 38.8 fs at room temperature and an adiabatic shift of the transition energy induced by the pump pulse, as identified by the smooth and continuous evolution of the macroscopic coherence through the dynamic shift.

Monolayer $WS_2$ has a direct gap at the K/K' symmetry points in the Brillouin zone, and a large exciton binding energy, ~320 meV[35]. Due to its two constituent elements and hexagonal lattice the crystal structure lacks inversion symmetry, which introduces significant Berry curvature around the K/K' valleys. Spin-orbit coupling splits the valence and conduction bands, which leads to a spin-valley locking of opposite sign. Consequently, circularly-polarized light selectively excites one of the two energetically-degenerate valleys, enabling so-called valleytronic applications[36]. This also leads to an optical Stark effect that is valley selective, when driven by a circularly polarized laser field.

To investigate the effect of ultrashort pulses in driving Floquet Bloch bands, we characterized the Stark shift in a monolayer of $WS_2$ on a $SiO_2$/Si substrate using a 25 fs probe pulse, resonant with the A-exciton energy (2.025 eV). The shift was induced by the sub-bandgap (1.850 eV) pump pulse whose duration was varied from 34 to 65 fs by altering the spectral bandwidth. The dynamics associated with the observed shift were determined by varying the delay between the pump and probe pulses. Details are provided in the Supplementary Information (SI).

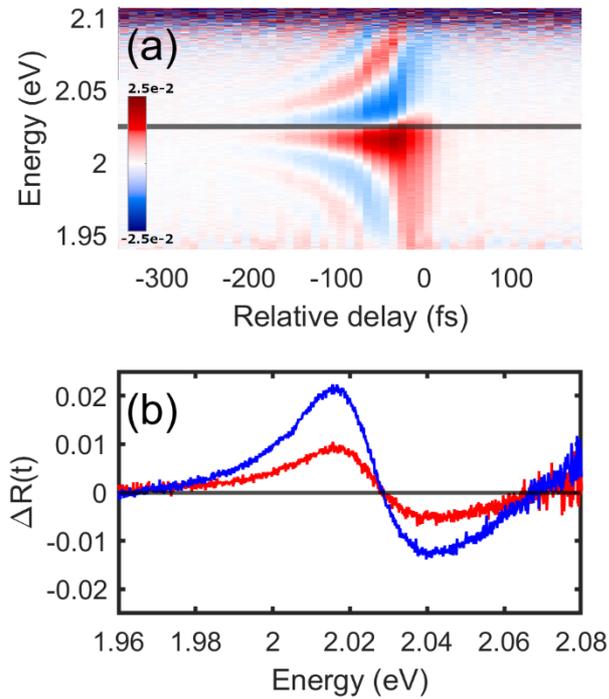

*Figure 1: (a) Measured transient reflectance from 34 fs, 50 µJ/cm$^2$ pump for co-circularly polarized probe; **(b)** Maximum signal at pulse overlap (relative delay $t = 0$) for 20 µJ/cm$^2$ (red) and 50 µJ/cm$^2$ (blue) pump fluence. Probe fluence was 3.5 µJ/cm$^2$.*

The equilibrium optical properties of the exfoliated WS$_2$ monolayer were characterized by spatially-resolved reflectance contrast and photoluminescence measurements performed at room temperature, as reported previously[37], and are shown in supplementary Fig. S1. To account for the influence of the intervening 300 nm SiO$_2$ layer and Si substrate, a 4-Lorentzian model was used to extract the equilibrium permittivity of the WS$_2$ layer using the transfer matrix method. The resulting equilibrium reflectance and absorption spectra, and the complex permittivity of the monolayer, are presented in Fig. S2. This enabled the conversion of the measured transient reflectance data (dR/R) to the transient absorption of the WS$_2$ monolayer (See SI for details).

The transient reflectance measured with co-circularly polarized pump and probe pulses is shown in **Error! Reference source not found.**(a) for a 34 fs pump pulse duration. The spectral response when pump and probe overlap in time ($t = 0$) is highlighted in Fig. 1(b), and shows a positive peak (increased reflectance due to decreased absorption) on the low-energy side of the exciton and a negative peak (decreased reflectance due to increased absorption) on the high-energy side. This dispersive profile arises from the transition being blue-shifted, as a result of the pump-induced optical Stark effect, and is consistent with previous observations[21,22].

The magnitude of the observed shift at $t = 0$ is 0.68 meV, as determined by the spectral weight transfer (SWT) calculated from the transient absorption spectrum[22,32] (See SI for details).

For relative delays $t > 0$ (i.e. for the pump pulse arriving first) there is no signal beyond the pulse overlap regime, confirming the expectation that the pump pulse is not generating any significant exciton population. Conversely, for relative delays $t < 0$ a response with fringes is observed [see Fig. 1(a)], persisting beyond -100 fs, well beyond the pulse overlap regime. The fringes represent a spectral interferogram arising from the interference between two pulses separated in time: the residual probe pulse, and a coherent response generated when the driving field (pump beam) is present.

In this case, the probe beam arrives first and excites a coherent superposition between ground and A-exciton states. This leads to a macroscopic polarization in the monolayer that remains in phase with the initial laser pulse for a time limited by decoherence (due to the finite homogeneous linewidth) and dephasing (due to the finite inhomogeneous linewidth). This macroscopic coherence can re-radiate as coherent photoluminescence, in the same direction as the probe, with a $\pi$ phase shift. This process, also known as free-induction decay, occurs whenever there is coherent resonant excitation. In the absence of the pump this coherent emission adds to the probe, giving the pump-free probe spectrum.

When the driving field is turned on (i.e. when the pump arrives) the energy separation between the ground and A-exciton state increases due to the optical Stark shift. If the eigenstates of the system evolve adiabatically, then the phase of the coherent superposition will evolve smoothly and the macroscopic polarization and the coherent emission will persist, only shifted to this higher energy for the duration of the driving field (pump pulse).

The shift will result in a decrease in the emission at the equilibrium energy and an increase at the blue-shifted energy. Because this shifted emission is $\pi$ out of phase with the incident probe, the net effect of its interference with the probe will be an increase in the measured intensity of the reflected probe on the red side and decrease on the blue side. This is precisely what is seen at pulse overlap ($t = 0$): the spectrally resolved signal gives the dispersive lineshape shown in Fig. 1(b). As the delay between the pump and probe increases, the time delay between the shifted coherent emission and the probe with which it interferes also increases, resulting in a spectral interferogram that changes with delay, as shown in Fig. 1(a). To observe this response, the macroscopic polarization must remain coherent, and thus the decay of the response with increasing delay yields a measure of the dephasing of the system.

To confirm that the spectral interference patterns do indeed correspond to a delay between pulses that matches the pump-probe delay, we Fourier transformed the data with respect to the energy (frequency) axis [Fig. 2(a)]. The result shows a signal that does shift in time by an amount equal to the relative delay.

The fringes of the interferogram can be removed by multiplying the complex spectral response by $e^{-i((\omega-\omega_0)(t+\tau))}$ (refer to SI), which reveals the intrinsic change in reflectance as a function of delay. The dispersive profile, previously only evident at $t = 0$, is now revealed as a function of relative delay between pump and probe in Fig. 2(b). From this, the transient absorption [Fig. S5] can be determined after correcting for substrate reflections and the equilibrium reflectance of the system. The peak of the dispersive feature appears at slightly negative times as it represents the convolution of the pump pulse with the exponential free-induction decay initiated by the arrival of the probe.

Previous work studying the optical Stark shift in GaAs/AlGaAs quantum wells observed similar results at low temperature. In that system, however, the coherence times are on the order of picoseconds, and relatively long (200 fs/150 fs) pulses were used[30–32]. The fringes seen here for $t < 0$ are also similar to what is observed in perturbed free-induction decay, where the free induction is effectively stopped by the pump pulse. The fringes in that case can arise from non-adiabatic processes and are symmetric about the peak of the absorption. This is in contrast to the observations here, where the fringes are antisymmetric about the equilibrium peak absorption due to the adiabatic spectral shift of the coherent emission.

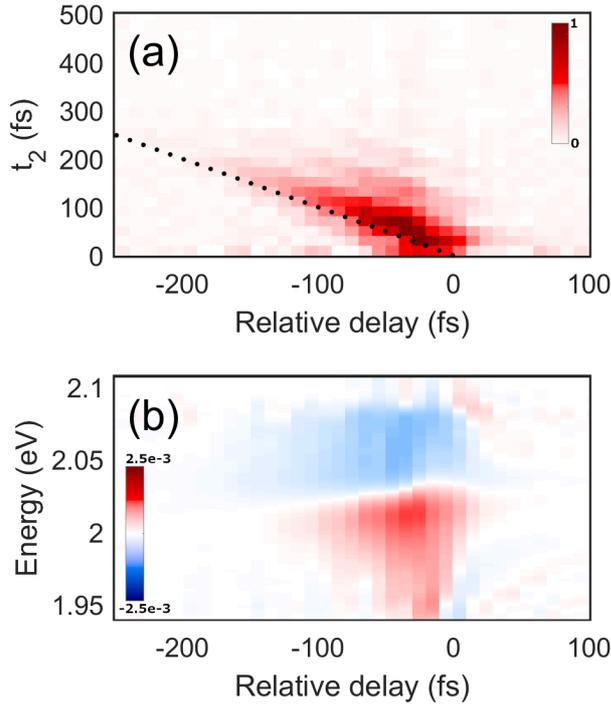

*Figure 2: (a) The Fourier-transform of the measured response. The line of slope -1 highlights that the shift in time of the transient feature is equal to the relative delay between pulses. (b) The transient change in reflectance, ΔR(t), of the WS$_2$ monolayer after correction for substrate effects.*

To confirm the origin of these fringes we performed simple simulations based on the Liouville von Neumann equation, assuming that the transition energy is adiabatically shifted, that is, the macroscopic coherence is maintained and evolves smoothly as the shift is induced, with the magnitude of the spectral shift following a Gaussian profile matching the pump pulse (refer to SI for details). Figs. 3 and S6 shows the result of these simulations, which qualitatively match the corresponding experimental measurement [Fig. 1(a)]. While these simulations are only to first order in the electric field, and the spectral shift is only introduced phenomenologically, their qualitative agreement with experimental data supports the proposed mechanism behind the observed interferogram and the adiabatic evolution of the eigenstates. In contrast, simulations involving either a smooth or rapid reduction of the coherence caused by the pump lead to a symmetric fringe pattern [Fig. S7], which is starkly different to our results.

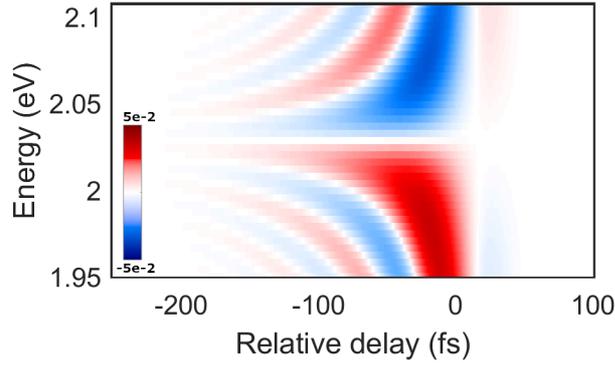

*Figure 3: Simulated pump-probe differential reflectance signal for a 34 fs pump pulse, reproduces the dispersive spectral response and fringes at negative delays present in the experimental data in Fig 1(a).*

The presence of these fringes therefore indicates that the optical Stark shift is switched on adiabatically, i.e., the energy of the state changes smoothly and maintains the macroscopic coherence of the system. The dynamics of the Stark shift can in principle then be determined from the calculated transient absorption [Fig. S5]. However, the response measured is a convolution of the dynamics of the shift induced by the pump pulse and the probe used to measure it, which includes both the pulse and the resultant free-induction decay. To extract these dynamics, we integrate the absolute value of the Fourier transformed data in Fig. 2(a), which gives the overall signal response amplitude. Fig. 4(a) shows the integrated signal amplitude as a function of delay for the 34 fs pump pulse. It is clear from this plot that at $t < 0$ the signal decays exponentially due to the dephasing, while for positive delays the signal appears to decay with the pulse overlap. This response was fit with an exponential convolved with a Gaussian, where the exponential describes the dephasing of the macroscopic coherence, while the Gaussian part represents the convolution of the probe pulse and the pump-induced Stark shift. From this fit, a value of 38 fs (+4/-3 fs) was obtained for the dephasing time, which is consistent with the linewidth of 32 meV FWHM measured in the linear differential reflectance spectrum with a Gaussian fit to the A exciton. To approximate the region of the $WS_2$ flake sampled by the probe pulse the linewidth was determined from a spatially-averaged region across the flake. It is worth noting that these measures are inclusive of inhomogeneous broadening due to random variations in the dielectric environment[38], and the homogeneous linewidth is likely narrower, as seen in room temperature coherent four-wave mixing measurements[39]. Still, the observation of a room temperature dephasing time of >38 fs speaks to the high-quality of this exfoliated $WS_2$ flake (without encapsulation), but also of the potential for using these materials in applications where room temperature coherence is important[40-42].

For positive delays ($t > 0$), the measured response appears to closely follow the pump pulse envelope. Since the signal amplitude is directly proportional to the magnitude of the Stark shift, this suggests that the shift varies with the instantaneous intensity. To examine this, we repeated these measurements for six different pump pulse durations by reducing the spectral bandwidth, thereby increasing the pump duration from 34 to 65 fs. To maximize time resolution the probe pulse duration was kept at 25 fs. Measurements were collected at fluences of 20 and 50 μJ/cm² for both co- and cross-circularly polarized pump and probe. Figs. S9 and S10 present the fitted dynamics of all pulses, along with the cross-correlation of pump and probe pulses, aligned to facilitate comparison. Fig. 4(b) shows the extracted Gaussian FWHM and exponential decay time for all pulses. The exponential decay time stayed constant, regardless of pulse duration or fluence, with a mean value of 39 ± 3 fs. The Gaussian FWHM, however, increased with the pulse duration, consistent with the convolution of the probe pulse and the Stark shift following the instantaneous intensity of the pump pulse.

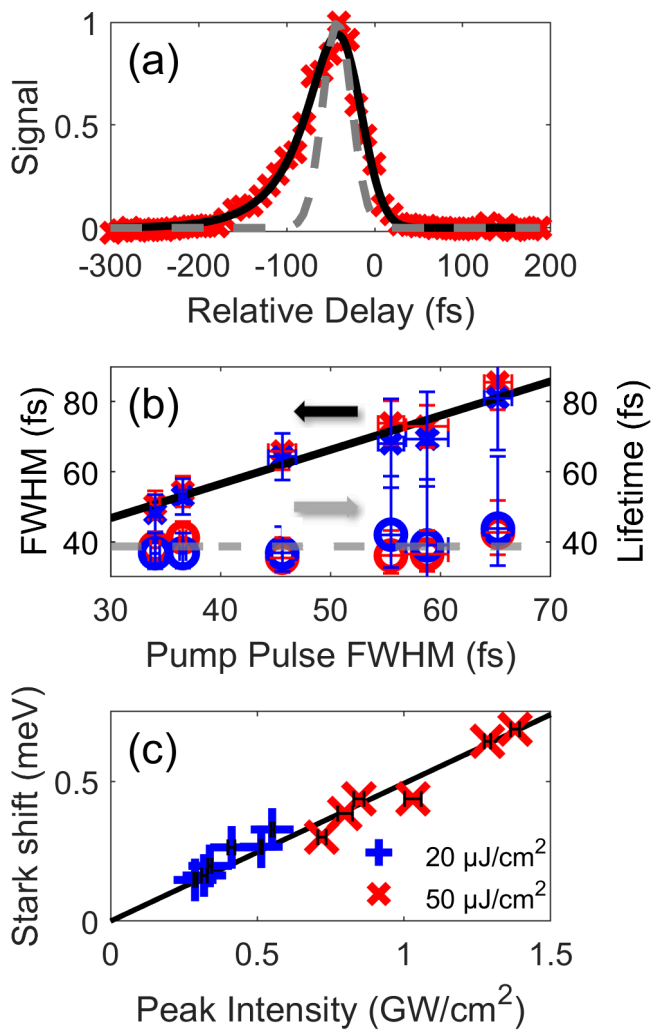

*Figure 4: (a) Integrated response of the absolute value of the Fourier transformed probe signal (red crosses), with fit line (black). The cross-correlation of the pulses (grey dashed line) is aligned to aid comparison; (b) FWHM of the Gaussian component (crosses) and rise time of the exponential component (circles) from the fits for different pulse durations and intensity (red – 50 µJ/cm$^2$; blue – 20 µJ/cm$^2$). A linear fit for the FWHM is included (y=0.9762\*x+17.4564). (c) Maximum excitonic Stark shift as a function of peak intensity for different pump pulse durations, together with a linear fit line (y=0.4931\*x; R-squared = 0.9497).*

Finally, we plot the maximum shift as a function of peak pump intensity for the different pulse durations and energies (Fig. 4(c)). If there is any non-adiabaticity, deviation from the behaviour predicted by Floquet theory, or some delay in the Stark shift turning on, the measured shift would deviate from the expected[21,22] linear trend at short pulse durations. However, for all pulse durations and both fluences, all data points lie on the same linear trendline, further confirming that the magnitude of the Stark shift follows the instantaneous intensity and that these ultrafast pulses drive the optical Stark effect as if the periodic perturbation was a continuous, monochromatic drive.

In conclusion, we have investigated the behavior of the valley-selective optical Stark effect in WS$_2$ monolayers. By controlling the spectral bandwidth and hence duration and peak intensity of the perturbing pump pulse, we have shown that even for pulses as short as 34 fs (~15 optical cycles) the optical Stark effect remains proportional to $|\mathbf{E}|^2$, and follows the instantaneous intensity. More significantly, the observed coherent interference provides strong evidence that the bands of the WS$_2$ monolayer are adiabatically shifted by the pump. This shift closely follows the envelope of the perturbing pulse, implying an instantaneous response.

Together, these findings provide compelling evidence that despite the broad spectral bandwidth and short duration of the pump pulse, the sample responds instantaneously while retaining its macroscopic coherence. This should be generally applicable to any material system and provide a basis from which to investigate quantum phase transitions.

See Supplemental Material at [URL will be inserted by publisher] for methods, sample details and equilibrium properties, details of the analyses and simulations discussed in the main text and the full set of raw experimental measurements. All files related to a published paper are stored as a single deposit and assigned a Supplemental Material URL. This URL appears in the article's reference list.

**Acknowledgements:**

This work was funded by the Australian Research Council centre of Excellence for Future Low-Energy Electronics Technologies (CE170100039). SKE would like to acknowledge Edbert Sie for helpful discussions in relation to the conversion of transient reflectance spectra to transient absorption.